\documentclass[a4 paper, 11pt,twoside]{article}
\usepackage{fullpage}                           % Makes the page margins smaller to a predefined one.
\usepackage[lmargin=0.6in,rmargin=0.6in,tmargin=0.6in,headsep=.2in]{geometry}
\usepackage{graphicx}
\usepackage{forloop}
\usepackage{etoolbox}
\usepackage{calc}
\usepackage{wrapfig,lipsum,booktabs} % To produce the empty space on the left
\usepackage{xtab} % To produce the empty space on the left
\usepackage[dvipsnames]{xcolor}
\usepackage[normalem]{ulem}
\usepackage{sectsty}% http://ctan.org/pkg/sectsty
\makeatletter
\def\@seccntformat#1{\@ifundefined{#1@cntformat}%
{\csname the#1\endcsname\;}%  default
{\csname #1@cntformat\endcsname}% individual control
}
\def\section@cntformat{\thesection.\;} % Dot after the section number
\def\subsection@cntformat{\thesubsection.\;} % Dot after the subsection number
\makeatother
\usepackage{hyperref}
\hypersetup{
    colorlinks=true,
    urlcolor=blue,
    linkcolor=black,
    citecolor = black}

\usepackage{fancyhdr}
\fancypagestyle{first}{%
  \fancyhf{}% clear all header and footer fields
}
\usepackage{amsmath,amsfonts,amssymb,amsthm,epsfig,epstopdf,url,array}
 \usepackage[retainorgcmds]{IEEEtrantools}
 \usepackage{makeidx,epsfig,lscape}
 \usepackage{xcolor,pict2e}
\usepackage{upgreek}
\newcommand{\dd}{\textnormal{\,d}}

\newcommand{\hh}{\textnormal{H\textsubscript{I}}}
\theoremstyle{definition}

\begin{document}
\thispagestyle{first}
\vspace*{3cm}
%%%%%%%%%  TITLE %%%%%%%%%%%%%%%%%
{\noindent\huge\bf Stability and Damping in the Disks of Massive Galaxies}\\[1cm]
%%%%%%%%%%%%%%%%  Author Data %%%%%%%%%%%%%%%%%%%
{\bf\large Marr John H.}\\[0.5cm]
%Unit of Computational Science, Building 250 \\
%Babraham Research Campus, Cambridge, CB22 3AT, UK \\
Unit of Computational Science \\
Hundon, Suffolk, CO10 8HD, UK \\
Email: john.marr@2from.com \\

%%%%%%%%%%%%%%%  The Abstract and Keywords 
{\color{Black}\rule{0.7\textwidth}{2pt}}\\[0.2cm]
{\color{Black}\bf\large Abstract}\\
After their initial formation, disk galaxies are observed to be rotationally stable over periods of $>6$~Gyr, implying that any large velocity disturbances of stars and gas clouds are damped rapidly on the timescale of their rotation. 
However, it is also known that despite this damping, there must be a degree of random local motion to stabilize the orbits against degenerate collapse. 
A mechanism for such damping is proposed by a combination of inter-stellar gravitational interactions, and interactions with the Oort clouds and exo-Oort objects associated with each star. 
Analysis of the gravitational interactions between two stars is a three-body problem, because the stars are also in orbit round the large virtual mass of the galaxy.
These mechanisms may produce rapid damping of large perturbations within a time period that is short on the scale of observational look-back time, but long on the scale of the disk rotational period for stars with small perturbations. 
This mechanism may also account for the locally observed mean perturbations in the Milky Way of 8-15~km/s for younger stars and 20-30~km/s for older stars.
%[151 words]
\vspace{0.5cm}\\
{\color{Black}\bf\large Keywords}\\
galaxies: kinematics and dynamics; spiral; lognormal density distribution; stability
\vspace{0cm}\\
{\color{Black}\rule{0.7\textwidth}{2pt}}

%%%%%%%%%%%%%%%%%  The Document Starts Here %%%%%%%%%%%%%%
\section{Introduction}
\label{secn:Intro}
General analysis of the gravitational stability of highly flattened, 'cold', massive stellar systems suggests that if they are assumed to be initially in approximate equilibrium between their self-gravitational and centrifugal forces, with purely circular motions and no random proper motions, then they are unstable to any density fluctuations \cite{2008gady.book.....B, 2000ARep...44..711T}. 
Such disks tend to form massive condensations within their own plane (the Jeans instability) unless their constituents have a minimum level of random motion as 'warm' or 'hot' disks in the directions parallel to the disk plane to stabilize them by migrating from overdense regions before a collapse can occur \cite{1929uau..book.....J, 2016ARep...60..116T}. 
Toomre \cite{1964ApJ...139.1217T} showed that the minimum root mean square (RMS) velocity dispersion required to suppress these axi-symmetric instabilities was $3.36G\mu/\kappa$~km/s (the Toomre stability criterion), where $G$ is the gravitational constant, and $\mu$ and $\kappa$ are the local values of the projected stellar density and the epicyclic frequency, respectively. 
This minimum was estimated to be $\sim20-35$~km~sec$^{-1}$ in the solar neighborhood of our Galaxy, a range which has since been matched by observations \cite{1964ApJ...139.1217T, 2018Natur.561..360A, 2021ApJ...910..163D}.

Disk galaxies comprise three major components: a central bulge of stars that harbour a massive black hole; a halo of dust and stars whose visible content is negligible compared to the disk; and a thin disk surrounding the bulge, that contains the majority of detectable mass in the system and contributes the major part of the total light and angular momentum \cite{2008gady.book.....B, 1958ApJ...128..465D, 1967PASJ...19..427T}. 
These galaxies are thought to be rotationally stable, with only slow changes in overall brightness over the past 6 Gyr \cite{2000ARep...44..711T, 2010AJ....140..663G}.
Mutual interactions by neighbouring stars will occur constantly, perturbing individual motions in complex ways, and may lead to gross instability unless there is some mechanism for damping. This paper uses numerical analysis to consider possible damping mechanisms for the fluctuations from stability in a massive disk galaxy. 

Perturbation methods traditionally start with an exact solution to a simplified form of the original problem, which in gravitational theory is typically a Keplerian ellipse, but the solution is exactly correct only when there are just two gravitating bodies. 
The introduction of a third body, or any non-Newtonian field such as the gravitational interaction using formulations from General relativity or the complex field of a galactic disk, does not yield a simplified form and in these cases numerical analysis may be used to describe perturbations from a stable state \cite{1973ApJ...186..467O, 1977ApJ...213..497H}.

Standard perturbation theory considers the disturbance from a stationary state, with such perturbations generally kept small to allow first order approximations to be made.
In this paper, we build a density model for the disk of a massive galaxy using M31 as a model, and use numerical analysis to investigate how large random motions---such as those arising from close interactions---might be rapidly damped to small random oscillations about their circular trajectory of $\sim20-30$~km~sec$^{-1}$ within the timescale of orbital rotation, and these mechanisms may be sufficient to stabilize the orbits against random gravitational fluctuations while retaining the stability demanded by Toomre \cite{1964ApJ...139.1217T}.

%%%%%%%%%%%%%%%%%%%%%%%%%%%%%%%%%%%%%%%%%%%%%%
\section{Building the surface density model}
\label{The model}
Because the stars are in orbit round the large mass of the galaxy, gravitational interactions between two stars is a three-body problem not amenable to simple mathematical analysis, but requiring iterative numerical analysis.
Modelling the gravitational motion of displacement of an individual disk star when subject to a perturbation requires a model for the gravitational potential throughout the disk.
The galaxy selected for modelling was the Messier 31 galaxy, as its size and proximity have enabled many detailed observations of its rotation curve (RC) and the surface density of its baryonic components \cite{1970ApJ...159..379R, 2006ApJ...641L.109C, 2010AA...511A..89C}.
Rotational velocities are measured as a bulk average over many stars in each disk area sampled in the surveys. 
A fundamental assumption is that tracers such as $\hh$ or H$\alpha$ in the disk move in circular orbits; non-circular motions due to collapsing gas clouds in star formation processes, bars, spiral density waves, and warps in the disk, cause distortions to the RC, but high-resolution data from surveys such as THINGS enable the effects of random non-circular motions to be minimized with the construction of a ``bulk'' velocity field showing the underlying undisturbed rotation \cite{2008AJ....136.2761O}.

Measuring the variation in Doppler shift across the disk enables its rotational velocity to be plotted as a function of radius from the galactic centre, allowing a rotational curve to be constructed. 
Although many RCs are relatively flat over much of their extent, there is generally a wider variety of curves than simple flatness, and modelling the mass-distribution to generate these curves has given rise to a number of models \cite{2016JMPh....7..680C, 2016JMPh....7.2177C, 2018Galax...6..115C, 2018P&SS..152...68H}, many of which invoke the presence of a dark matter (DM) halo \cite{2008AJ....136.2761O}.  
The observational distribution of surface brightness and $\hh$ density across the galactic disk approximates to an exponential form, and this has often been taken as representative of the underlying baryonic components of the disk.
Such a distribution however is poorly justified on theoretical grounds, and there have been a number of attempts to describe better or more accurate models \cite{2016JMPh....7..680C, 2016JMPh....7.2177C, 2018Galax...6..115C, 2018P&SS..152...68H}.
For example, considering the disk as a relaxed, stable assemblage of particles in an otherwise isolated system justifies stating that---except for the conserved macroscopic variables of total mass, energy and angular momentum---such a system has lost all information about its prior unrelaxed state and may consequently be considered as a system of maximum entropy with a lognormal (LN) distribution, and it has been shown that the rotation curves for a wide range of galactic disks are generally well described by a LN density distribution curve \cite{2015MNRAS.448.3229M, 2015MNRAS.453.2214M}.

\begin{figure}
   \centering
	\includegraphics[width=0.75\textwidth]{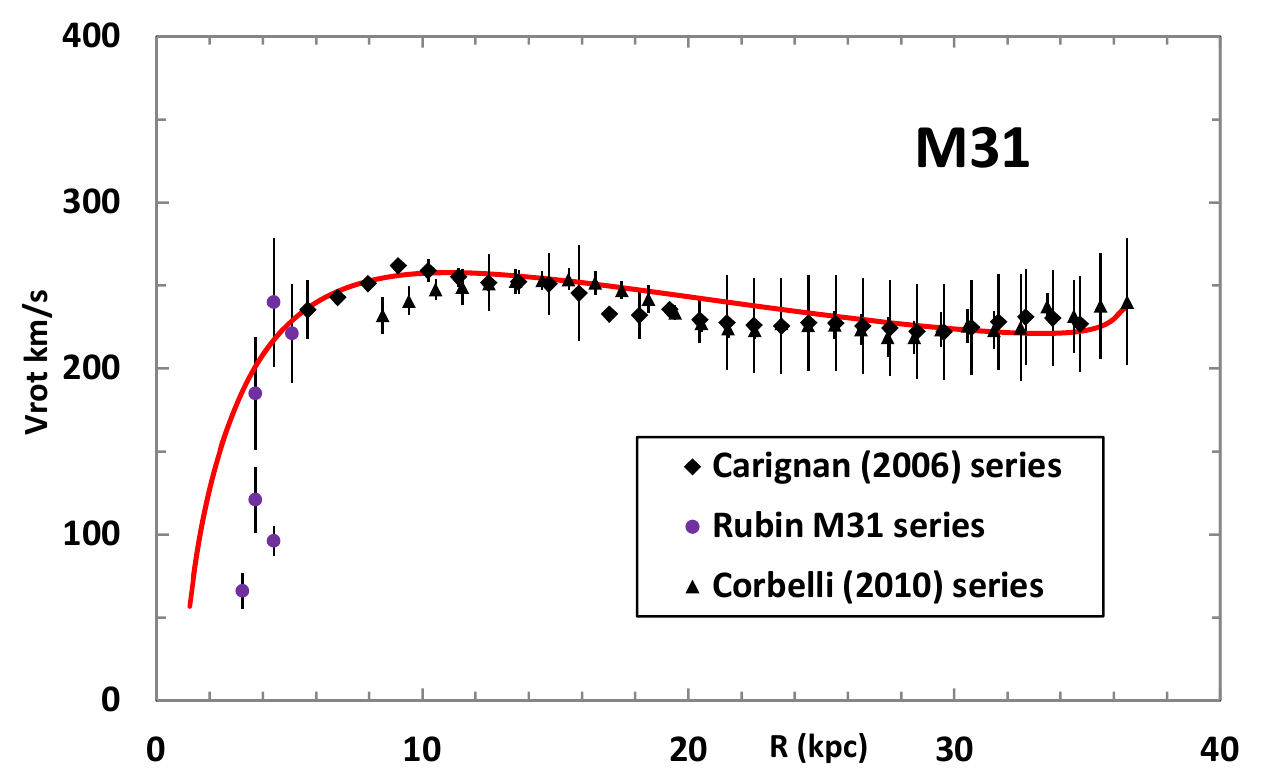}
   \caption{The rotational curve for M31. Overlain is the theoretical curve for a lognormal distribution of surface density with all gravitational mass confined to the disk, Newtonian gravitation, and assuming no DM halo. The best-fit for a lognormal density distribution is overlain (red line). Source data: \cite{1970ApJ...159..379R, 2006ApJ...641L.109C, 2010AA...511A..89C} } 
   \label{fig:M31}
	\vspace*{8pt}
\end{figure}

A typical disk has $\sim 10^{8}-10^{12}$ stars, and for large N it is usual to consider the average, statistical properties of the system rather than individual orbits. 
Although a dynamical approach to describe the relaxation process is difficult, especially as no exact description of the initial state is known, the self-similarity and stability of disk galaxies allows them to be considered as idealized equilibrium systems \cite{2020Galax...8...12M}. 
Because disks are thin compared with their radius, most analytical studies assume them to have negligible thickness and describe them in terms of a pure surface density function, $\Sigma(r)$.
The lognormal surface density for such a system is described by a distribution with three principle parameters: 
a characteristic radial scale length $r_\mu$ (kpc); 
the logarithmic standard deviation of the radius $\sigma$ ; 
and a characteristic surface density parameter $\Sigma_0$ ($M_\odot$~kpc$^{-2}$) \cite{2015MNRAS.448.3229M}. 
The variety of curves generated by a LN distribution are consistent with the three broad classes of observational RCs described by Verheijen \cite{2001ApJ...563..694V}, with some rising towards their termination, some reaching a plateau, and others reaching a peak before declining again. 
The LN model provides a curve consistent with the original observations of M31 by Rubin and Ford \cite{1970ApJ...159..379R}, Carignan $et~al$ \cite{2006ApJ...641L.109C}, and Corbelli $et~al$  \cite{2010AA...511A..89C}, and gives a convenient method of generating $\Sigma(r)$ for numerical stability analysis (Figure~\ref{fig:M31}). 
Furthermore, the general shape of the velocity curve for M31 is broadly similar to recent curves for the Milky Way Galaxy \cite{2017A&A...597A..39M} (Figure~1).

For the galactic disk, the general mathematical form for a lognormal distribution may be modified to the more physical form (Eq.~\ref{eq:LN2}):
\begin{equation}
\Sigma(r)=\frac{\Sigma_0}{(r/r_\mu)\sigma\sqrt{2\pi}}\exp{\left(-\frac{[\log(r/r_\mu)]^2}{2\sigma^2}\right)}\,,
\label{eq:LN2}
\end{equation}
where $\Sigma(r)$ is the disk surface density ($M_\odot$~kpc$^{-2}$) and $r$ is the radial variable (kpc). 

Unlike the exponential distribution, the lognormal distribution matches the expected probability distribution for disk systems. 
The radius where the stars orbit must be $>0$; the distribution is highly skewed rather than Gaussian; normalisation of the function to unity yields the total probability that an individual star is certain to be somewhere in the disk; and it is smoothly asymptotic to zero at the core where rotation is unsupported, rather than peaking to a cusp --- characteristics that satisfy the observed mass-density distribution of the disks of spiral galaxies  \cite{2012msma.book.....F}.
For a given density distribution, numerical integration allows the rotation curve to be derived, and can include a bulge or any boundary conditions at $R_{max}$. 

Although Eq.~\ref{eq:LN2} is exact only in the limit $r\rightarrow\infty$, in practice $\Sigma(r)\rightarrow0$ as $r\rightarrow R_{max}$, which is the maximum radius for observations (kpc) beyond which gas and dust at the galactic periphery become undetectable. 
The fact that $\Sigma(r)\rightarrow0$ as $r\rightarrow0$ is also reasonable, reflecting the collapse of the rotation curve where bulge stars predominate near the galactic center. 
RMS error-minimisation curve-fitting was used to generate a best-fit output rotation curve for M31 \cite{2015MNRAS.448.3229M}.
With $R_{max}=36.5$~kpc, these parameters were: $\Sigma_0=7.21\times10^9$~$M_\odot$~kpc$^{-2}$; $r_\mu=4.5$~kpc; $\sigma=1.15$ to generate the curve of Figure \ref{fig:M31}. 
Integration of Eq.~\ref{eq:LN2} gave the theoretical total mass (stars, gas and dust) of the M31 disk as $2.96\times10^{11}$~$M_\odot$, which lies between an observational stellar mass of $1.03\times10^{11}$~$M_\odot$ \cite{2015IAUS..311...82S} and more recent observations of $1.3-1.6\times10^{12}$~$M_\odot$, using the motion of external satellite galaxies \cite{2010MNRAS.406..264W}.
M31 has a relatively flat curve and the derived LN curve fits the observations reasonably well. 
%%--------------------
%%--------------------

The theoretical LN distribution extends to infinite radius which is clearly unphysical, and in Figure~\ref{fig:M31} the surface density was stopped abruptly at $R_{max}$, producing a terminal rise beyond 32~kpc.
This is a feature of any RC with an abrupt termination \cite{2015MNRAS.448.3229M}, but also one seen in the data of Corbelli $et~al$ \cite{2010AA...511A..89C}. 
In practice, the terminal density will fall away more gradually, but observationally it is difficult to detect this termination because any observable matter will already be included in the disk, and matter beyond the detectable disk boundary will by definition be unobserved. 
Nevertheless, many observers have reported that their $\hh$ observations showed no evidence of stopping at their limit of detection, and---with the increasing sensitivity of observations---there is now evidence for some $\hh$ and molecular gas components extending beyond the original disk boundaries, usually described by adding further exponential components to the disk boundary as a biaxial or triaxial disk \cite{2008AJ....135...20E, 2013AJ....146..104H}.

%%%%%%%%%%%%%%%%%%%%%%%%%%%%%%%%%%%%%%%%%%%%%%%%%
\section{Stability of the disk}
\label{sec:Stability}
There are two components to the motion of a galactic star: a) its velocity of rotation as part of the flat rotation curve, and b) pseudo-random perturbations from interactions with other stars or massive dust clouds.
Peebles \cite{1969ApJ...155..393P} conjectured that galactic spins originated from induced tidal torques from neighbouring structures with in-falling gas subsequently forming the disk.
Although the subsequent history of disk formation is uncertain, it is clear that some mechanism was in place to distribute the initial range of velocities into the regular quasi-circular orbits now observed.
In the absence of external forces, this mechanism must have involved some form of damping to bring about an approximate equipartition of velocities in each orbit, and to dampen induced oscillations.

%%--------------------
\begin{figure}[ht]
   \centering
	\includegraphics[width=\textwidth]{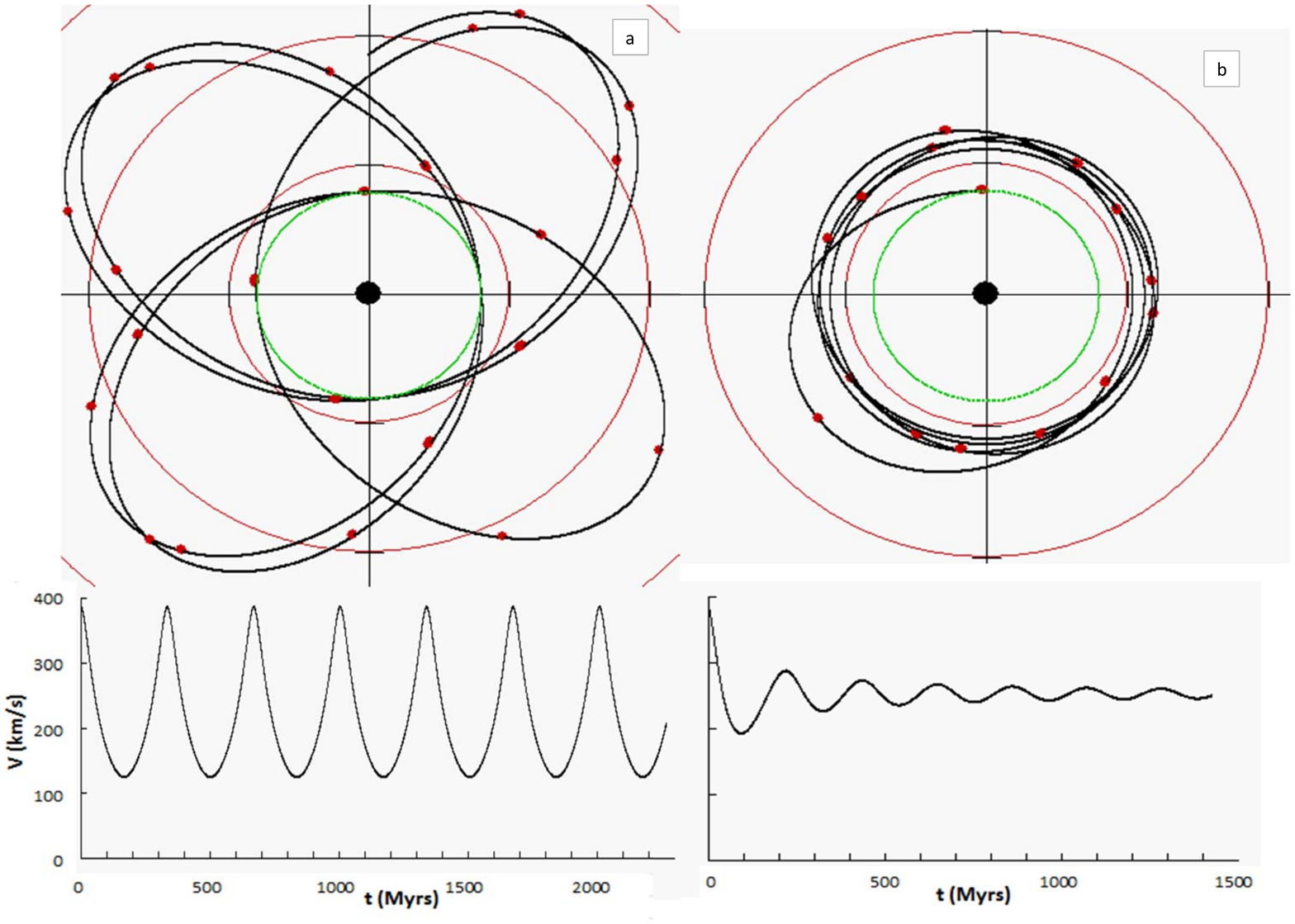}
   \caption{A star of $1~M_\odot$ orbiting at 8~kpc from the galactic centre with an initial stable orbital velocity of 259.2~km~s$^{-1}$ (green circle), boosted to $\times1.5$ its initial velocity.
   (a) undamped. (b) Damped: $\zeta_0= 0.0001$~km~s$^{-1}$~Myr$^{-1}$ per km~s$^{-1}$, with a relaxation time of $\sim500$~Myr {(\it{see text})}. 
   Times for 5 orbits are shown, marked with red dots every 100~Myr. moving anticlockwise viewed from above. Red circles = 10~kpc scale markers.   
   The bottom two graphs show the undamped and damped orbital velocities (km~sec$^{-1}$) $vs.$ time. 
   %Scale markers are: 100~km~s$^{-1}$ (vertical) and 100~Myr (horizontal). 
   } 
   \label{fig:orbits}
	\vspace*{8pt}
\end{figure}
%%--------------------
%%--------------------
\begin{figure}
   \centering
	\includegraphics[width=0.75\textwidth]{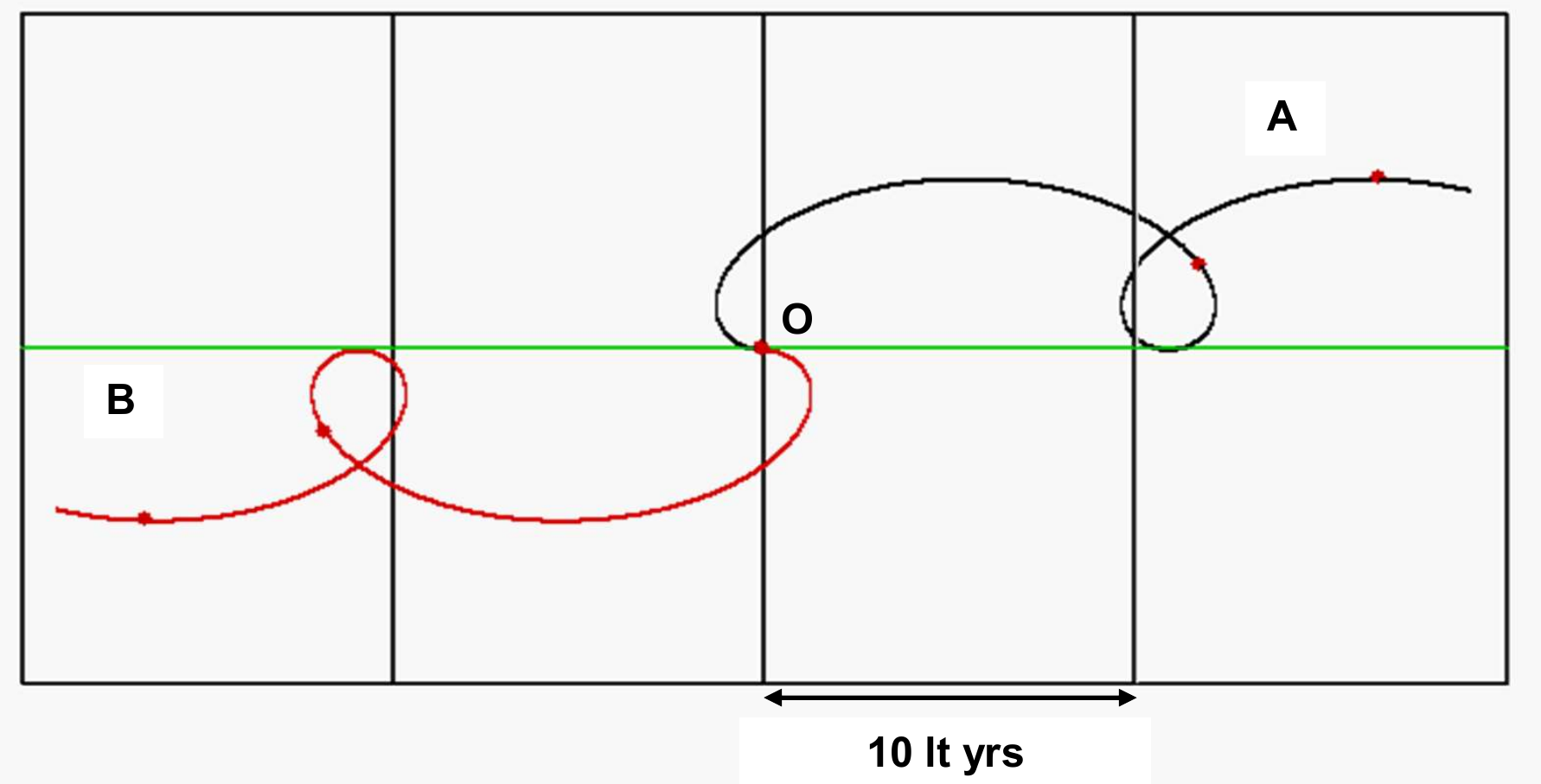}
   \caption{A magnified frame centred on a stable star of $1~M_\odot$ initially in a stable orbit at 8~kpc (centre O on green line) from the galactic centre with an initial stable orbital velocity of 259.2~km~s$^{-1}$, anticlockwise as viewed from above. 
   (A) Boosted by +0.01\%.  (black curve). 
   (B) As (A), but with the stable orbital velocity decreased by 0.01\% (red curve).
   Frame grid spacing 10~lt.~yrs. 
  Time markers (red dots) every 100~Myr. } 
   \label{fig:increase}
	\vspace*{8pt}
\end{figure}
%%--------------------

Figure~\ref{fig:orbits} illustrates the motion of a typical star of mass $M_\odot$ orbiting at a radial distance of 8~kpc from the centre (green circle), with and a radial velocity of 259.2~km~sec$^{-1}$ for its stable circular motion.
The orbit was computed assuming the gravitational potential for M31 is of the form proposed in Section~\ref{The model}.
The radial distance markers (red circles) are at 10~kpc intervals, and orbital time markers (red dots) are shown every 100~Myrs.
Figure~\ref{fig:orbits}(a) illustrates the consequent undamped orbit for the star displaced from its initially circular 'equilibrium' state by a massive boost of $\times1.5$ to the radial velocity, computed by numerical analysis, while Figure~\ref{fig:orbits}(b) shows the same boost, but damped by a factor $\zeta_0$, defined more fully in Section~\ref{section: oort}. 
The nature of the damping mechanism is unspecified here, but is given an arbitrary value $\zeta_0=0.0001$ (km~s$^{-1}$ per Myr per km~s$^{-1}$), and is assumed to be a function of the volume swept out in a given time interval, and hence of the relative velocity of the star ($v$) to the field through which it is moving at any instant.

Although at the time of formation large displacements such as those shown in Figure~\ref{fig:orbits} may have been common, they are probably rare during the mature phase of the galaxy's existence when random displacements by encounters with other star systems are likely to be small on the scale of the radial motion.
These large motions are, however, in concurrence with the motions noted by Michtchenko $et~al$ \cite{2018ApJ...863L..37M} who measured the proper velocities of 3,105,498 stars from the second data set of the Gaia mission with good distance estimates within 1~kpc of the Sun, then converting the positions, parallaxes, proper motions on the sky, and radial velocities of the stars into Cartesian Galactic phase-space positions and velocities.
Their Gaia DR2 measures showed large scale corkscrewing that appeared to be sustained by resonant features of the spiral arms to maintain the spiralling motion.

Figure~\ref{fig:increase} represents a highly magnified view of Figure~\ref{fig:orbits}, centred on a star orbiting at 8~kpc ('O' on the green line) from the galactic centre, with the stable position of the unperturbed star in the centre (red dot). 
The frame dimensions are only 10 light years per square, hence at this scale ($\sim 26,000\times$ the scale of Figure~\ref{fig:orbits}), the grid lines appear linear as the curvature of the radial lines is too small to be apparent.
The frame is co-rotating anti-clockwise to the centre as seen from above, i.e. right to left.
Any star rotating in a stable orbit on the green 8~kpc line will not appear to move; stars closer in will appear to move to the left (overtaking), while stars further out will appear to move to the right (be overtaken).
The central star is then subjected to a small boost (Figure~\ref{fig:increase}, A) or decrease (Figure~\ref{fig:increase}, B) in its orbital velocity of $\pm 0.01\%$ respectively.
These are undamped motions, but in each case the star corkscrews away from its initial position because of the conservation of the new angular momentum as it orbits the galactic centre.

%%%%%%%%%%%%%%%%%%%%%%%%%%%%%%%%%%%%%%%%%%%
\section{Potential mechanisms for damping}
Three principle mechanisms are considered: a) damping by interactions between the Oort clouds of other stars during the motion of the displaced star; b) damping from interstellar dust and gas clouds; and c) gravitational interactions with other stars as the displaced star moves past them.
Perturbations to the equilibrium gravitational potential affect the averaged velocity distribution of stars \cite{1538-3881-137-3-3520} through participation in collective motions of all the particles in the system \cite{1972.book.....K, 1964ApJ...139.1217T}.
These collective processes lead to a quasi-chaotic path through velocity space, in the sense that a change in the initial parameters smaller than observational uncertainty will lead to a completely different path and set of interactions within a time-frame that is small compared to the orbital time. 
In practice, the number of stars in mutual proximity is highly variable and they move in the gravitational potential of the whole disk.
There are, for example, 12 stars within 10~lt~yrs of the Sun, each contributing to the overall motion of the Sun \cite{2014ApJ...786L..18L} with interactions also occurring in the $z-$plane, and the resultant interactions are therefore immensely more complex.   

%%%%%%%%%%%%%%%%%%%%%%%%%%%%%%%%%%%%%%%%%%%
\subsection{Damping by Oort clouds}
\label{section: oort}
The size of a system of mass $M$ without a sharp boundary may be characterised by its gravitational sphere of influence, or the extent of its Hill sphere, defined here as the radius, $r_g$, within which small individual masses may orbit without escaping \cite{2008gady.book.....B}. 
For a star of stellar mass $1~M_\odot$ such as the sun, the Oort cloud is thought to extend to approximately $1.5\times10^{13}$~km, and this value is taken to approximate the star's gravitational sphere of influence, $r_g$.
Such a star system may be expected to sweep out a volume of $\sim7.07\times10^{26}\times v_{drift}$~km$^3$sec$^{-1}$, where $v_{drift}$~km~sec$^{-1}$ is the relative difference in motion between the displaced star and the stable orbital velocity at that radius.
The masses of the Oort clouds surrounding such systems are unknown, and the density and mass of the background population of exo-Oort cloud objects is also unknown \cite{2018ApJ...866..131M}, therefore we may only estimate possible values, extrapolating from the limited information available for the Oort cloud of the solar system \cite{1983AA...118...90W}. 
This may contain $10^{11}-10^{12}$ icy bodies, with a total estimated mass of $10^{25}-10^{26}$~kg and a mean density $\sim2\times10^{-15}$~kg~km$^{-3}$, although in one estimate it may approach 2\% of the solar mass, i.e. $\sim4\times10^{28}$~kg \cite{1986EM&P...36..187M}.

We assume a majority of stars to be moving in quasi-stable circular orbits with their associated Oort clouds, through which a displaced star of mass $M$ moves with an initial drift velocity $v_0$~km~s$^{-1}$, relative to the locally stable velocity frame of circulation. 
The migrating star exchanges mass at a rate $\Delta m /\Delta t$ as it traverses the local star systems, such that its velocity changes at the rate $\Delta v/\Delta t$, partitioned between the drifting star system and its remaining clouds. 
In contrast to the multiple accumulating impulses of the migrating star, the interacting orbiting star systems with initially zero velocity relative to the rotational frame each gains an exchange in mass $\Delta m$ as individual impulses, and their individual changes in velocity are therefore neglected.
We assign a mean density and radial extent to the cloud system of $\rho_g$~kg~km$^{-3}$ and $r_g$~km respectively, and state: 
\begin{equation}
    \Delta m/ {\Delta t}=\rho_g A v~\textnormal{kg~s$^{-1}$}\,,
    \label{eq:m}
\end{equation}
where $\Delta m/\Delta t$ is the incremental time increase in mass $M$ at time $t$, $A=2\pi r_g^2$ is the swept out area, and $v$ is the drift velocity through the cloud.  

Eq.~\ref{eq:m} may be rewritten as:
\begin{equation}
    \Delta m/{\Delta t}=2M\zeta_0 v ~~~\textnormal{kg~s$^{-1}$}\,.
    \label{eq:m2}
\end{equation}

Here $\zeta_0=C_t \rho_g A/2M$ is defined as the damping coefficient for the Oort cloud,  and $C_t=3.154\times10^{13}$~sec/Myr is the dimensionless conversion factor.
By conservation of kinetic energy:
\begin{equation}
    \frac{1}{2}Mv^2=\frac{1}{2}(M+\Delta m)(v+\Delta v)^2\,,
    \label{eq: KE}
\end{equation}
\begin{equation}
    \textnormal{or~~~~}\frac{\Delta v}{v} \approx \frac{\Delta m}{2M}\,,
    \label{eq:dv2}
\end{equation}
where $\Delta v $ is the incremental change in initial velocity, $v$, associated with acquiring the mass $\Delta m$. 
Substituting $\Delta m$ from Eq.~\ref{eq:m2} in Eq.~\ref{eq:dv2} and neglecting terms in $\Delta v^2$ and $\Delta v \Delta t$\,,
\begin{equation}
    \Delta v/\Delta t \approx \zeta_0 v^2 \textnormal{~km~s$^{-1}$ Myr$^{-1}$}\,.
    \label{eq: dv}
\end{equation}

Then letting $\Delta v\rightarrow \dd v$, $\Delta t\rightarrow \dd t$ and integrating with limits $v=v_0$ at $t=0$:
\begin{equation}
    v\approx \frac{v_0}{1+\zeta_0 v_0 t}\,.
    \label{eq: v}
\end{equation}

Assuming a mean stellar separation of $4\times10^{13}$~km ($4.2$ lt yrs), then the damping factor for Figure~\ref{fig:orbits}(b) is $\zeta_0\approx 0.0001$~km~sec$^{-1}$Myr$^{-1}$ per km~sec$^{-1}$ and this value was selected for these figures to demonstrate the overall effect of smooth damping, irrespective of its cause.
The motion is not that of a simple harmonic oscillator and the decay $\dd v/\dd t$ of Eq.~\ref{eq: dv} is a power function of the velocity typical of dynamical damping with rapid initial decay but much slower late damping, rather than an exponential form. 
The damping half-life is $t_{1/2}\sim(\zeta_0 v_0)^{-1}$ from an initial velocity, $v_0$, in contrast to the fixed half-life of an exponential decay.
For high initial relative velocities, the displaced motion rapidly settles to a new quasi-circular orbit, with a relaxation time $t_{1/2}\sim77$~Myrs for the high displacement velocity in this example.
However, at long time intervals there remains a residual oscillating proper motion which is slow to dissipate, with $t_{1/2}\sim1$~Gyr for a displacement velocity $v_0=\pm10$~km~s$^{-1}$ and the given value of $\zeta_0$.
For a star moving slower than the local field, it may be noted that the same mechanism would cause an increase in velocity towards that of the field velocity but with the same damping coefficient.

%%%%%%%%%%%%%%%%%%%%%%%%%%%%%%%%%%%%%%%%%%%
\subsection{Interactional damping by interstellar dust and gas clouds}
\label{section: gas}
The interstellar medium (ISM) is a mixture of gas and dust remaining from: a) the formation of the galaxy; b) ejection by stars; and c) accretion from outside the galaxy. 
Observations of the apex and anti-apex direction of the Sun moving relative to the local ISM (LISM) confirm that the orientation of the flux direction of interstellar matter into the solar system as determined by relative velocities is almost parallel to the ecliptic plane \cite{1993AdSpR..13..121W}.

The gas is very diffuse: at its densest in the plane of the Galaxy the particle number density is 10$^{12}$ to 10$^{18}$ atomic nuclei km$^{-3}$, with some in the form of single neutral atoms, some in the form of simple molecules, and some existing as ions.
Its chemical composition is about 91\% hydrogen, 9\% helium.
It is observationally important because spectroscopic emission lines from the gas enable measurements of the mass and dynamics of the gas, including rotation curves.
Mass measurements of $\hh$ mass are generally multiplied by a factor 1.4 to take into account the presence of Helium, but dust and molecular and ionized gas are not quantified.
The mass of atomic hydrogen as $1.67\times10^{-27}$~kg, with a total gas density varying between $10^{-15}$ and $2.3\times10^{-9}$~kg~km$^{-3}$. 

The total density of dust in the ISM is thought to be considerably less than the gas density, and Draine $et~al$ \cite{2007ApJ...663..866D} suggest that $M_{dust}/M_{(HI+H2)}\approx0.01$. 
The composition of the dust particles is highly variable, and grains may vary in size by a factor of $100:1$, with the detection of larger grains supporting collision models for particle growth in the ISM.

The lack of small particles in the measured mass distribution from impact measurements compared to ISM conditions is a result of their depletion by solar mechanisms. 
Three methods of decelerating interaction may exist for a migrating star system: 
repulsion by the star's magnetic field deflecting particles with mass $m<10^{-17}$ kg within the heliosphere;
radiation pressure repulsion which is particularly important for particles with masses $10^{-17}< m < 10^{-16}$ kg, while for larger particles ($m>10^{-15}$ kg), gravitational focusing may aid their capture \cite{2000JGR...10510317M}.
The capture area for ISM gas and dust (Eq.~\ref{eq:m}) may be much lower than that of the Oort cloud if confined to the periheliosphere out to the heliopause beyond the termination shock, with $A\sim2\times10^{11}$~km$^2$. 
For this small area, $\zeta_g<3\times10^{-15}$ and is negligible unless we postulate interactions from the area of the gravitational radius to bear an influence on ISM damping, but this is not considered further here.

%%%%%%%%%%%%%%%%%%%%%%%%%%%%%%%%%%%%%%%%%%%%%%%%%
\subsection{Damping by interstellar interactions}
\label{interstellar}
Although the motions and interactions of individual stars are quasi-chaotic, it is possible to compute an approximation for possible damping by interstellar interactions. Consider a star of mass $M$ moving through a star field with a differential velocity $v$.
Let $\Delta p$ be the mean exchange in angular momentum/interaction; the star will then have a mean number of interactions of $v/s$ per Myr, where $s$ is the mean separation between stars, so the mean rate of change of momentum per encounter per Myr is:
\begin{equation}
    \frac{v\Delta p}{s} = \frac{M~v\Delta v}{s}\textnormal{~Myr$^{-1}$}.
\end{equation}

We may again define an interstellar damping coefficient $\xi_S$ such that 
\begin{equation}
    \xi_S=\frac{\Delta v}{Ms} \textnormal{~~km s$^{-1}$ Myr$^{-1}$ per km s$^{-1}$ $M_\odot^{-1}$}\,.
\end{equation}

For a star of 1~$M_\odot$ and taking a mean stellar separation of 4.2~lt~yrs, similar damping to $\xi_0$ for the Oort clouds may require $\xi_S=0.000126$, or $\Delta v = 0.000126v$ km~s$^{-1}$ per encounter  (Section~\ref{section: oort}).
If the star is moving more slowly than the field, this becomes an increase in velocity towards the mean field velocity.
%%--------------------
\begin{figure}
   \centering
	\includegraphics[width=0.6\textwidth]{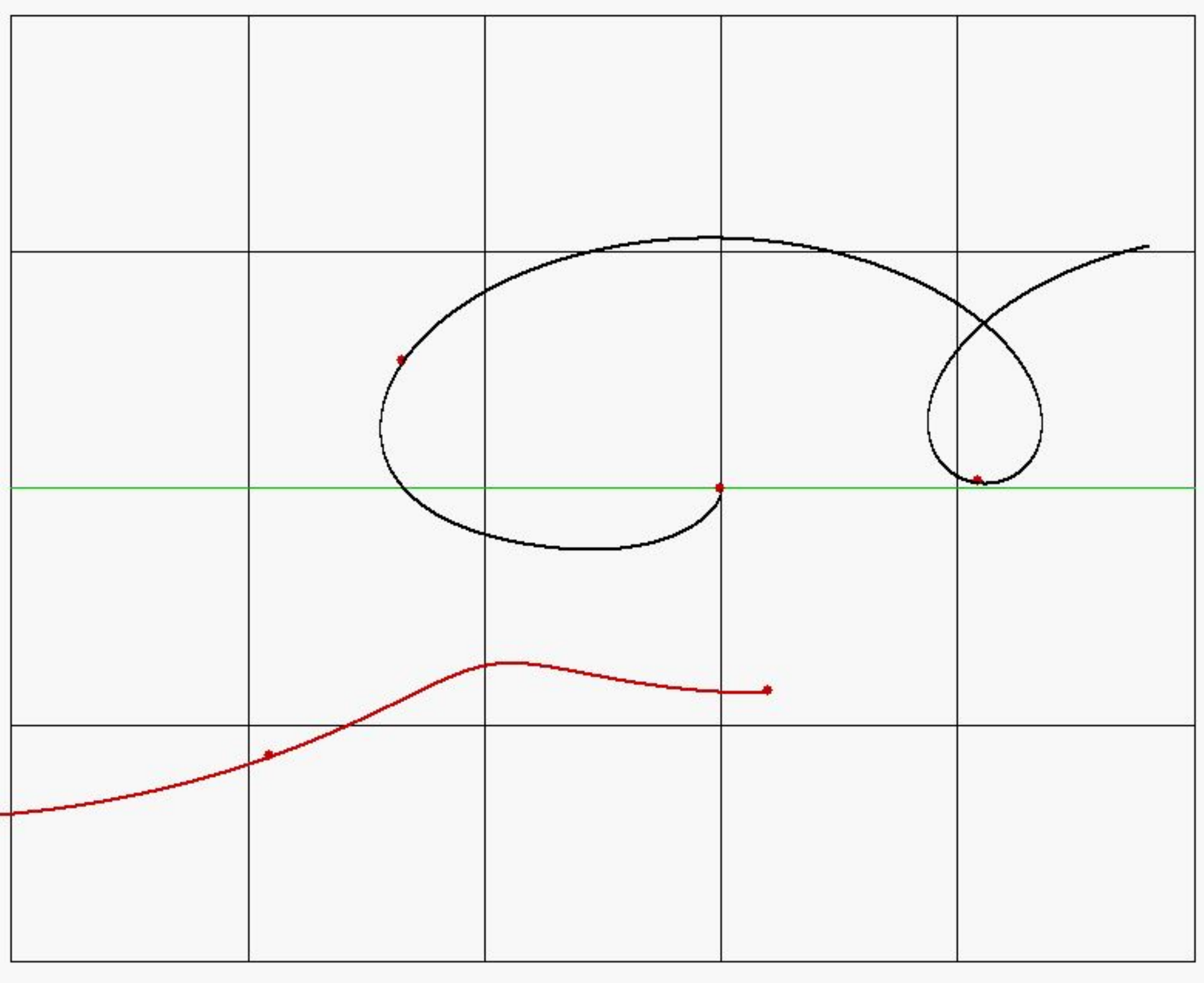}
   \caption{Perturbation of a reference star of $1~M_\odot$ (black orbit) orbiting the galactic centre at 8~kpc under mutual gravity from a second star of $2~M_\odot$ (red orbit) closer to the centre, 'overtaking' from the right with an initial separation of 8.6~lt yrs from the reference star. Frame grid spacing $=10$~lt.~yrs. Time markers (red dots) are every 100~Myr. Closest separation is 5.45 light years.} 
   \label{fig:sirius}
	\vspace*{8pt}
\end{figure}
%%--------------------

Figure~\ref{fig:sirius} illustrates the mutual motion between a reference star of $1~M_\odot$ and a second star of mass $2~M_\odot$ in the same plane, both orbiting at approximately identical radii (8~kpc and 7.99736~kpc respectively) and with similar rotational velocities (259.1791~km/s and 259.1777~km/s respectively).
The second star is in a galactic orbit 8.6 light years closer to the galactic centre than the lighter star, and their relative motion results in the heavier star 'overtaking' the lighter star as it moves from right to left in Figure~\ref{fig:sirius}.
Although initially moving on adjacent orbits with similar tangential velocities, the 'stationary' star gains angular momentum and after a close encounter begins to spiral in a higher orbit (black curve) with a boost of $+0.018\%$ in a similar manner to the star given a boost in Figure~\ref{fig:increase}.
The gravitational pull is slight, but it is cumulative and sufficient to pull the lighter star significantly from its orbit before it spirals away.   
The heavier star loses angular momentum and its orbit moves closer towards the centre in a similar manner to that illustrated in Figure~\ref{fig:increase} (B).
Because of its mass, the heavier star has approximately twice the AM of the lighter star but after the interaction the final AM of the overtaking heavier star is reduced and the final AM of the lighter star is increased by a corresponding amount, with conservation of the total combined AM throughout the interaction. 

%%%%%%%%%%%%%%%%%%%%%%%%%%%%%%%%%%%%%%%%%%
\subsection{Motion in the z-plane}
A number of recent observations have indicated that some stars have sufficient momentum to escape the disk completely. 
The perturbation illustrated in Figure~\ref{fig:sirius} is relatively small, but an interaction between a massive and a low mass star moving in close proximity may induce a considerable displacement in the smaller star.
Observations within our own Galaxy suggest there are less than 1 in 10,000 high-speed stars capable of contributing to the halo in the $z$-plane \cite{2018ApJ...868...25B}, and Bromley $et~al$ \cite{2018ApJ...868...25B} identified just over 100 high-speed stars from $\sim$1.5 million nearby stars (10-15~kpc) in the Gaia DR2 archive with measured parallax, proper motion, and radial velocity, of which only two are likely to be hyper-runaway stars rather than just bound outliers, with a nearly 100\% chance of being unbound.
These observations, coupled with the quasi-circular orbits for the vast majority of stars in the disk, lend support to a damping mechanism to preserve the thin disk over the many Gyr of look-back observations of disks.

For the majority of stars above the galactic plain, it is postulated that similar mechanisms for damping occur to those described for rotational damping, although the damping coefficients may be correspondingly smaller because of the low density of stars in the halo compared with the disk.
Several models for the vertical mass distribution in the disk have been proposed, such as a model of the form proposed by van der Kruit \cite{1988A&A...192..117V}:
\begin{equation}
    \varrho(z)=2^{-2/n}\varrho_e \textnormal{sech}^{2/n}(nz/2z_e)\,, 
\end{equation}
where $n=1$ corresponds to the isothermal distribution $\varrho(z)=(\varrho_e/4) \textnormal{sech}^2(z/2z_e)$, while $n=\infty$ corresponds to the exponential model $\varrho(z)=\varrho_e \exp{(-z/z_e)}$ \cite{1988A&A...192..117V}.

Although detailed motion in the $z-$plane will differ according to the model, perturbations in the absence of damping will result in oscillations about the midline of the disk in all these models, with a corresponding $v_{drift}$ in the $z-$plane.

%%%%%%%%%%%%%%%%%%%%%%%%%%%%%%%%%%%%%%%%%%%%%%%%%
\subsection{Local perturbations in the disk}
\label{section: interstellar_motion}
Superimposed on the bulk rotational velocity, individual stars have a peculiar velocity with a Gaussian distribution along each coordinate in velocity space---the Schwarzschild distribution---with a low velocity dispersion in the range 8--15~km s$^{-1}$ \cite{2008gady.book.....B} for stars with ages $10^8$--$10^9$~yr, and a higher dispersion range of 25--50~km s$^{-1}$ for older stars, although it may be noted that this relationship was based in relatively few stars and from older data.
These peculiar velocities arise from mutual collisionless interactions with other stars that take the complete system toward thermal quasi-equilibrium, with a collective relaxation timescale of $t_{relax} \simeq$~2--3~Gyr locally in the Milky Way \cite{1538-3881-137-3-3520}.

Differences in population numbers as a function of mass may be considered as an imbalance between the rates of star formation and their lifetimes. 
The formation rate is generally approximated by the Initial Mass Distribution Function (IMF), taken to be of the Salpeter form (Eq.~\ref{eq:Salpeter}):
\begin{equation}
\label{eq:Salpeter}
    \Phi(m)\Delta m \propto \left(\frac{m}{M_\odot}\right)^{-\alpha}
\end{equation}

With $\alpha=2.35$ to a first approximation, and for a star formation rate $\Re$ stars/Myr/unit volume over time $\Delta t$, there will be $\Re V\Phi(m)\Delta m\Delta t$ stars born with mass between $m$ and $m+\Delta m$, with a preponderance of low mass stars.
This will be balanced by the the star extinction rate, a function of the expected lifetime, $\tau_{M}$, of any specific star on the main sequence.
Although not accurate for very massive or very light stars due to its use of a single value mass-luminosity relationship for main sequence stars, to a first approximation this is an inverse function of the mass $m$ of that star  (Eq.~\ref{eq:time_mass}):
\begin{equation}
\label{eq:time_mass}
    \frac{\tau_{M}}{\tau_\odot}\simeq\left(\frac{m}{M_\odot}\right)^{-2.5}\,,
\end{equation}
where $M_\odot$ and $\tau_\odot$ are the mass and lifetime of the Sun respectively \cite{1988asco.book.....H}.
Therefore the overall number $N$ with mass between $m$ and $m+\Delta m$ after time $t$ is of the form:
\begin{equation}\label{eq:Sk}
\left\{ \begin{array}{lcl}
N\propto \Re V\left({m}/{M_\odot}\right)^{-2.35}\Delta m\,t & \mbox{for population } & \tau_M>t \\
N\propto \Re V\left({m}/{M_\odot}\right)^{-2.35}\Delta m\,\tau_M \mbox{ or }\propto \Re V\left(m/M_\odot\right)^{-4.85}\Delta m & \mbox{for population } & \tau_M<t \,.
\end{array}\right.
\end{equation}

Hence within any given observational volume $V$, there is an abundance of long-lived low mass stars, but few short-lived high mass stars, although this may be less apparent within regions of rapid star formation such as the spiral arms and the immediate stellar neighbourhood, with a preponderance of younger stars.
Although it has been argued that this gradual increase in peculiar velocity with time is by an unknown process increasing the velocity dispersion after the birth of the stars \cite{1538-3881-137-3-3520}, we propose that this is a natural consequence of the inverse variation in the mean age of a star with its mass.
Because high mass stars will be disrupted to a far lesser degree than low mass stars, this lends support to the observation that stars younger than 1~Gyr will have a lower range of velocity dispersion than older stars  \cite{2008gady.book.....B}. 

\section{Discussion}
Two notable features of many disk galaxies are the general similarity of their structure, with a well-defined disk of gas and stars rotating in approximately circular orbits about a massive centre, and the stability of the disks over the period of several Gyrs for which observations are possible \cite{2000ARep...44..711T, 2010AJ....140..663G}, with one recent observation confirming an early spiral galaxy at a redshift of $z=4.2603$ \cite{2020Natur.581..269N}. 
This remarkable stability seems to persist from the earliest appearance of disk galaxies despite strong interaction perturbations from adjacent stars and the presence of inhomogeneities such as a bar and spiral arms within the disk. 
The instability of local 'cold' regions is thought to be sufficient to allow spiral arms to develop, while the increasing density within these arms may prevent runaway instability by a process of negative feedback \cite{2008gady.book.....B}.
This suggests that there is some mechanism to damp out the inherent instabilities and ensure that the bulk motions remain circular and confined to the narrow disk, rather than degenerating into some form of elliptical galaxy.
This paper considers possible damping mechanisms with the potential to restore these perturbations asymptotically towards the local field velocity.

Velocity perturbations are defined here as any excess or deficient velocity differential relative to the local 'bulk' velocity of the rotating annuli in the disk.
In addition to short-range Newtonian gravitational interactions between neighbouring stars, there is a gravitational potential from the whole disk that is responsible for the bulk motions. 
The gravitational acceleration from the Milky Way disk at the position of the sun ($\sim 8$~kpc from the centre) is $a_{disk}\sim 2.17\times 10^{-10}$~m~s$^{-2}$, in contrast to that from a star of 1~solar mass at a distance of 4~light years from the sun of $a_{star}\sim 9.27\times10^{-14}$~m~s$^{-2}$, and this central acceleration must therefore be included in the numerical analysis of stellar interactions. 
This makes the prediction of any stellar-stellar interaction a three-body problem, which in turn demands a model for the gravitational potential throughout the disk.

There have been many attempts to model the gravitational potential for disk galaxies  \cite{2008AJ....136.2761O, 2016JMPh....7..680C, 2016JMPh....7.2177C, 2018Galax...6..115C, 2018P&SS..152...68H, 2015MNRAS.448.3229M}, including a simple exponential model.
Over the relatively short range of inter-stellar interactions however, the local gravitational potential from the bulk disk will not show much variation and the exact model chosen is therefore not critical.
The model selected for the analysis in this paper is the lognormal density distribution model because of its simplicity and generally good fit to a wide range of galaxies \cite{2015MNRAS.448.3229M}.
The galaxy chosen for analysis is M31, both because of its similarity to the Milky Way and because its proximity has resulted in many good measurements of the rotational velocity fields \cite{1970ApJ...159..379R, 2006ApJ...641L.109C, 2010AA...511A..89C}.

The age of the halo of the Milky Way is 13.5~Gyr \cite{2007ApJ...660L.117F} but that of the disk is only $8.8\pm 1.7$~Gyr \cite{2005A&A...440.1153D}, suggesting that damping from any disruption at its initial formation and subsequent internal perturbations occurred within this time scale.
The model described here suggests that damping of fluctuations may occur through mutual perturbations with other stars, by interactions with the Oort clouds, and exo-Oort cloud objects thought to be present round most stars. 
Possible mechanisms for this imply that a damping coefficient of $\zeta\approx 0.0001$~km~sec$^{-1}$Myr$^{-1}$ per km~sec$^{-1}$ will damp initial perturbations within  $\sim$0.5-1~Gyr, equivalent to $\sim2-3$ orbits for a star at 8kpc from the galactic centre.
Density requirements for damping by interstellar dust and gas suggest that these are unlikely to have any important contribution to overall damping (Section~\ref{section: gas}), and if such damping came only from Oort clouds it would require a total cloud mass in excess of 2\% of the stellar mass, i.e. $\sim4\times10^{28}$~kg for a star of 1~$M_\odot$, with the whole mass distributing the change in angular momentum of the passing star (Section~\ref{section: oort}). 
Although there have been proposals that the density of Oort clouds and other debris may be high \cite{1986EM&P...36..187M}, it is more likely that only a small part of this will absorb the change in angular momentum required, and inter-stellar interactions provide the remainder of the interchange for overall damping.

Unlike the damping of simple harmonic motion, the damping of perturbed stars is not an exponential function of time, but results in low level damping for small differentials, with a residual perturbation of the order of $\sim20-30$~km~sec$^{-1}$ maintained by the differential rotation of neighbouring stars.
Such damping may be sufficient to stabilize the orbits against large fluctuations while retaining the small fluctuations required to maintain orbital stability demonstrated by Binney and Tremaine \cite{2008gady.book.....B} and Toomre \cite{1964ApJ...139.1217T}, and matches the perturbations described in our own galaxy from  observations with HIPPARCOS \cite{2005A&A...442..929A}.

%%%%%%%%%%%%%%%%%%%%%%%%%%%%%%%%%%%%%%%%%%
\vspace{6pt} 

%%%%%%%%%%%%%%%%%%%%%%%%%%%%%%%%%%%%%%%%%%
{\textbf{Acknowledgments}}

I wish to thank the anonymous reviewers for many valuable comments, and Alexander Tutukov for his discussions and insights into disk stability.

%%%%%%%%%%%%%%%%%%%%%%%%%%%%%%%%%%%%%%%%%%
{\textbf{Conflicts of Interest}}

The author declares no conflict of interest. 

%%%%%%%%%%%%%%%%%%%%%%%%%%%%%%%%%%%%%%%%%%
\bibliographystyle{sn-mathphys}
\bibliography{galaxies3}


\begin{thebibliography}{48}
\ifx \bisbn   \undefined \def \bisbn  #1{ISBN #1}\fi
\ifx \binits  \undefined \def \binits#1{#1}\fi
\ifx \bauthor  \undefined \def \bauthor#1{#1}\fi
\ifx \batitle  \undefined \def \batitle#1{#1}\fi
\ifx \bjtitle  \undefined \def \bjtitle#1{#1}\fi
\ifx \bvolume  \undefined \def \bvolume#1{\textbf{#1}}\fi
\ifx \byear  \undefined \def \byear#1{#1}\fi
\ifx \bissue  \undefined \def \bissue#1{#1}\fi
\ifx \bfpage  \undefined \def \bfpage#1{#1}\fi
\ifx \blpage  \undefined \def \blpage #1{#1}\fi
\ifx \burl  \undefined \def \burl#1{\textsf{#1}}\fi
\ifx \doiurl  \undefined \def \doiurl#1{\url{https://doi.org/#1}}\fi
\ifx \betal  \undefined \def \betal{\textit{et al.}}\fi
\ifx \binstitute  \undefined \def \binstitute#1{#1}\fi
\ifx \binstitutionaled  \undefined \def \binstitutionaled#1{#1}\fi
\ifx \bctitle  \undefined \def \bctitle#1{#1}\fi
\ifx \beditor  \undefined \def \beditor#1{#1}\fi
\ifx \bpublisher  \undefined \def \bpublisher#1{#1}\fi
\ifx \bbtitle  \undefined \def \bbtitle#1{#1}\fi
\ifx \bedition  \undefined \def \bedition#1{#1}\fi
\ifx \bseriesno  \undefined \def \bseriesno#1{#1}\fi
\ifx \blocation  \undefined \def \blocation#1{#1}\fi
\ifx \bsertitle  \undefined \def \bsertitle#1{#1}\fi
\ifx \bsnm \undefined \def \bsnm#1{#1}\fi
\ifx \bsuffix \undefined \def \bsuffix#1{#1}\fi
\ifx \bparticle \undefined \def \bparticle#1{#1}\fi
\ifx \barticle \undefined \def \barticle#1{#1}\fi
\ifx \bconfdate \undefined \def \bconfdate #1{#1}\fi
\ifx \botherref \undefined \def \botherref #1{#1}\fi
\ifx \url \undefined \def \url#1{\textsf{#1}}\fi
\ifx \bchapter \undefined \def \bchapter#1{#1}\fi
\ifx \bbook \undefined \def \bbook#1{#1}\fi
\ifx \bcomment \undefined \def \bcomment#1{#1}\fi
\ifx \oauthor \undefined \def \oauthor#1{#1}\fi
\ifx \citeauthoryear \undefined \def \citeauthoryear#1{#1}\fi
\ifx \endbibitem  \undefined \def \endbibitem {}\fi
\ifx \bconflocation  \undefined \def \bconflocation#1{#1}\fi
\ifx \arxivurl  \undefined \def \arxivurl#1{\textsf{#1}}\fi
\csname PreBibitemsHook\endcsname

%%% 1
\bibitem{2008gady.book.....B}
\begin{bbook}
\bauthor{\bsnm{{Binney}}, \binits{J.}},
\bauthor{\bsnm{{Tremaine}}, \binits{S.}}:
\bbtitle{{Galactic} {Dynamics}}.
\bpublisher{{Princeton University Press, Princeton, NJ USA.}}, \blocation{ }
(\byear{2008})
\end{bbook}
\endbibitem

%%% 2
\bibitem{2000ARep...44..711T}
\begin{barticle}
\bauthor{\bsnm{{Tutukov}}, \binits{A.V.}},
\bauthor{\bsnm{{Shustov}}, \binits{B.M.}},
\bauthor{\bsnm{{Wiebe}}, \binits{D.S.}}:
\batitle{{The Stellar Epoch in the Evolution of the Galaxy}}.
\bjtitle{Astronomy Reports}
\bvolume{44},
\bfpage{711}--\blpage{718}
(\byear{2000}).
\doiurl{10.1134/1.1320496}
\end{barticle}
\endbibitem

%%% 3
\bibitem{1929uau..book.....J}
\begin{bbook}
\bauthor{\bsnm{{Jeans}}, \binits{J.H.}}:
\bbtitle{{The} {universe} {around} {us}}.
\bpublisher{{Cambridge University Press}}, \blocation{ }
(\byear{1929})
\end{bbook}
\endbibitem

%%% 4
\bibitem{2016ARep...60..116T}
\begin{barticle}
\bauthor{\bsnm{{Tutukov}}, \binits{A.V.}},
\bauthor{\bsnm{{Fedorova}}, \binits{A.V.}}:
\batitle{{Formation of ring structures in galactic disks during close passages
  of galaxies}}.
\bjtitle{Astronomy Reports}
\bvolume{60}(\bissue{1}),
\bfpage{116}--\blpage{128}
(\byear{2016}).
\doiurl{10.1134/S1063772915120082}
\end{barticle}
\endbibitem

%%% 5
\bibitem{1964ApJ...139.1217T}
\begin{barticle}
\bauthor{\bsnm{{Toomre}}, \binits{A.}}:
\batitle{{On the gravitational stability of a disk of stars.}}
\bjtitle{\apj}
\bvolume{139},
\bfpage{1217}--\blpage{1238}
(\byear{1964}).
\doiurl{10.1086/147861}
\end{barticle}
\endbibitem

%%% 6
\bibitem{2018Natur.561..360A}
\begin{barticle}
\bauthor{\bsnm{{Antoja}}, \binits{T.}},
\bauthor{\bsnm{{Helmi}}, \binits{A.}},
\bauthor{\bsnm{{Romero-G{\'o}mez}}, \binits{M.}},
\bauthor{\bsnm{{et.~al.}}}:
\batitle{{A dynamically young and perturbed Milky Way disk}}.
\bjtitle{\nat}
\bvolume{561}(\bissue{7723}),
\bfpage{360}--\blpage{362}
(\byear{2018})
{\href{https://arxiv.org/abs/1804.10196}{{arXiv:1804.10196}}}
{[astro-ph.GA]}.
\doiurl{10.1038/s41586-018-0510-7}
\end{barticle}
\endbibitem

%%% 7
\bibitem{2021ApJ...910..163D}
\begin{barticle}
\bauthor{\bsnm{{Das}}, \binits{I.}},
\bauthor{\bsnm{{Basu}}, \binits{S.}}:
\batitle{{Linear Stability Analysis of a Magnetic Rotating Disk with Ohmic
  Dissipation and Ambipolar Diffusion}}.
\bjtitle{\apj}
\bvolume{910}(\bissue{2}),
\bfpage{163}
(\byear{2021})
{\href{https://arxiv.org/abs/2011.08876}{{arXiv:2011.08876}}}
{[astro-ph.SR]}.
\doiurl{10.3847/1538-4357/abdb2c}
\end{barticle}
\endbibitem

%%% 8
\bibitem{1958ApJ...128..465D}
\begin{barticle}
\bauthor{\bsnm{{de Vaucouleurs}}, \binits{G.}}:
\batitle{{Photoelectric photometry of the Andromeda nebula in the UBV system}}.
\bjtitle{\apj}
\bvolume{128},
\bfpage{465}
(\byear{1958}).
\doiurl{10.1086/146564}
\end{barticle}
\endbibitem

%%% 9
\bibitem{1967PASJ...19..427T}
\begin{barticle}
\bauthor{\bsnm{{Takase}}, \binits{B.}}:
\batitle{{Distribution of Mass, Angular Momentum, and Rotational Energy in the
  Galaxy and NGC 224}}.
\bjtitle{\pasj}
\bvolume{19},
\bfpage{427}
(\byear{1967})
\end{barticle}
\endbibitem

%%% 10
\bibitem{2010AJ....140..663G}
\begin{barticle}
\bauthor{\bsnm{{Gurovich}}, \binits{S.}},
\bauthor{\bsnm{{Freeman}}, \binits{K.}},
\bauthor{\bsnm{{Jerjen}}, \binits{H.}},
\bauthor{\bsnm{{Staveley-Smith}}, \binits{L.}},
\bauthor{\bsnm{{Puerari}}, \binits{I.}}:
\batitle{{The Slope of the Baryonic Tully-Fisher Relation}}.
\bjtitle{\aj}
\bvolume{140},
\bfpage{663}--\blpage{676}
(\byear{2010})
{\href{https://arxiv.org/abs/1004.4365}{{arXiv:1004.4365}}}.
\doiurl{10.1088/0004-6256/140/3/663}
\end{barticle}
\endbibitem

%%% 11
\bibitem{1973ApJ...186..467O}
\begin{barticle}
\bauthor{\bsnm{{Ostriker}}, \binits{J.P.}},
\bauthor{\bsnm{{Peebles}}, \binits{P.J.E.}}:
\batitle{{A Numerical Study of the Stability of Flattened Galaxies: or, can
  Cold Galaxies Survive?}}
\bjtitle{\apj}
\bvolume{186},
\bfpage{467}--\blpage{480}
(\byear{1973}).
\doiurl{10.1086/152513}
\end{barticle}
\endbibitem

%%% 12
\bibitem{1977ApJ...213..497H}
\begin{barticle}
\bauthor{\bsnm{{Hunter}}, \binits{C.}}:
\batitle{{On Secular Stability, Secular Instability, and Points of Bifurcation
  of Rotating Gaseous Masses}}.
\bjtitle{\apj}
\bvolume{213},
\bfpage{497}--\blpage{517}
(\byear{1977}).
\doiurl{10.1086/155181}
\end{barticle}
\endbibitem

%%% 13
\bibitem{1970ApJ...159..379R}
\begin{barticle}
\bauthor{\bsnm{{Rubin}}, \binits{V.C.}},
\bauthor{\bsnm{{Ford}}, \binits{W.K.} \bsuffix{Jr.}}:
\batitle{{Rotation of the Andromeda Nebula from a Spectroscopic Survey of
  Emission Regions}}.
\bjtitle{\apj}
\bvolume{159},
\bfpage{379}
(\byear{1970}).
\doiurl{10.1086/150317}
\end{barticle}
\endbibitem

%%% 14
\bibitem{2006ApJ...641L.109C}
\begin{barticle}
\bauthor{\bsnm{{Carignan}}, \binits{C.}},
\bauthor{\bsnm{{Chemin}}, \binits{L.}},
\bauthor{\bsnm{{Huchtmeier}}, \binits{W.K.}},
\bauthor{\bsnm{{Lockman}}, \binits{F.J.}}:
\batitle{{The Extended H I Rotation Curve and Mass Distribution of M31}}.
\bjtitle{\apjl}
\bvolume{641},
\bfpage{109}--\blpage{112}
(\byear{2006})
{\href{https://arxiv.org/abs/astro-ph/0603143}{{astro-ph/0603143}}}.
\doiurl{10.1086/503869}
\end{barticle}
\endbibitem

%%% 15
\bibitem{2010AA...511A..89C}
\begin{barticle}
\bauthor{\bsnm{{Corbelli}}, \binits{E.}},
\bauthor{\bsnm{{Lorenzoni}}, \binits{S.}},
\bauthor{\bsnm{{Walterbos}}, \binits{R.}},
\bauthor{\bsnm{{Braun}}, \binits{R.}},
\bauthor{\bsnm{{Thilker}}, \binits{D.}}:
\batitle{{A wide-field H I mosaic of Messier 31. II. The disk warp, rotation,
  and the dark matter halo}}.
\bjtitle{\aap}
\bvolume{511},
\bfpage{89}
(\byear{2010})
{\href{https://arxiv.org/abs/0912.4133}{{arXiv:0912.4133}}}
{[astro-ph.CO]}.
\doiurl{10.1051/0004-6361/200913297}
\end{barticle}
\endbibitem

%%% 16
\bibitem{2008AJ....136.2761O}
\begin{barticle}
\bauthor{\bsnm{{Oh}}, \binits{S.-H.}},
\bauthor{\bsnm{{de Blok}}, \binits{W.J.G.}},
\bauthor{\bsnm{{Walter}}, \binits{F.}},
\bauthor{\bsnm{{Brinks}}, \binits{E.}},
\bauthor{\bsnm{{Kennicutt}}, \binits{R.C.} \bsuffix{Jr.}}:
\batitle{{High-Resolution Dark Matter Density Profiles of THINGS Dwarf
  Galaxies: Correcting for Noncircular Motions}}.
\bjtitle{\aj}
\bvolume{136},
\bfpage{2761}--\blpage{2781}
(\byear{2008})
{\href{https://arxiv.org/abs/0810.2119}{{arXiv:0810.2119}}}.
\doiurl{10.1088/0004-6256/136/6/2761}
\end{barticle}
\endbibitem

%%% 17
\bibitem{2016JMPh....7..680C}
\begin{barticle}
\bauthor{\bsnm{{Christodoulou}}, \binits{D.M.}},
\bauthor{\bsnm{{Kazanas}}, \binits{D.}}:
\batitle{{The Case against Dark Matter and Modified Gravity: Flat Rotation
  Curves Are a Rigorous Requirement in Rotating Self-Gravitating Newtonian
  Gaseous Discs}}.
\bjtitle{Journal of Modern Physics}
\bvolume{7},
\bfpage{680}--\blpage{698}
(\byear{2016})
{\href{https://arxiv.org/abs/1510.05534}{{arXiv:1510.05534}}}
{[astro-ph.GA]}.
\doiurl{10.4236/jmp.2016.77067}
\end{barticle}
\endbibitem

%%% 18
\bibitem{2016JMPh....7.2177C}
\begin{barticle}
\bauthor{\bsnm{{Christodoulou}}, \binits{D.M.}},
\bauthor{\bsnm{{Kazanas}}, \binits{D.}}:
\batitle{{Exact Axisymmetric Solutions of the 2-D Lane-Emden Equations with
  Rotation}}.
\bjtitle{Journal of Modern Physics}
\bvolume{7},
\bfpage{2177}--\blpage{2187}
(\byear{2016})
{\href{https://arxiv.org/abs/1701.03953}{{arXiv:1701.03953}}}
{[astro-ph.GA]}.
\doiurl{10.4236/jmp.2016.715189}
\end{barticle}
\endbibitem

%%% 19
\bibitem{2018Galax...6..115C}
\begin{barticle}
\bauthor{\bsnm{{Criss}}, \binits{R.}},
\bauthor{\bsnm{{Hofmeister}}, \binits{A.}}:
\batitle{{Galactic Density and Evolution Based on the Virial Theorem, Energy
  Minimization, and Conservation of Angular Momentum}}.
\bjtitle{Galaxies}
\bvolume{6},
\bfpage{115}
(\byear{2018}).
\doiurl{10.3390/galaxies6040115}
\end{barticle}
\endbibitem

%%% 20
\bibitem{2018P&SS..152...68H}
\begin{barticle}
\bauthor{\bsnm{{Hofmeister}}, \binits{A.M.}},
\bauthor{\bsnm{{Criss}}, \binits{R.E.}},
\bauthor{\bsnm{{Criss}}, \binits{E.M.}}:
\batitle{{Verified solutions for the gravitational attraction to an oblate
  spheroid: Implications for planet mass and satellite orbits}}.
\bjtitle{Planetary and Space Science}
\bvolume{152},
\bfpage{68}--\blpage{81}
(\byear{2018}).
\doiurl{10.1016/j.pss.2018.01.005}
\end{barticle}
\endbibitem

%%% 21
\bibitem{2015MNRAS.448.3229M}
\begin{barticle}
\bauthor{\bsnm{{Marr}}, \binits{J.H.}}:
\batitle{{Galaxy rotation curves with lognormal density distribution}}.
\bjtitle{\mnras}
\bvolume{448},
\bfpage{3229}--\blpage{3241}
(\byear{2015})
{\href{https://arxiv.org/abs/1502.02949}{{arXiv:1502.02949}}}.
\doiurl{10.1093/mnras/stv216}
\end{barticle}
\endbibitem

%%% 22
\bibitem{2015MNRAS.453.2214M}
\begin{barticle}
\bauthor{\bsnm{{Marr}}, \binits{J.H.}}:
\batitle{{Angular momentum of disc galaxies with a lognormal density
  distribution}}.
\bjtitle{\mnras}
\bvolume{453},
\bfpage{2214}--\blpage{2219}
(\byear{2015})
{\href{https://arxiv.org/abs/1507.04515}{{arXiv:1507.04515}}}.
\doiurl{10.1093/mnras/stv1734}
\end{barticle}
\endbibitem

%%% 23
\bibitem{2020Galax...8...12M}
\begin{barticle}
\bauthor{\bsnm{{Marr}}, \binits{J.H.}}:
\batitle{{Entropy and Mass Distribution in Disc Galaxies}}.
\bjtitle{Galaxies}
\bvolume{8}(\bissue{1}),
\bfpage{12}
(\byear{2020})
{\href{https://arxiv.org/abs/2002.03110}{{arXiv:2002.03110}}}
{[astro-ph.GA]}.
\doiurl{10.3390/galaxies8010012}
\end{barticle}
\endbibitem

%%% 24
\bibitem{2001ApJ...563..694V}
\begin{barticle}
\bauthor{\bsnm{{Verheijen}}, \binits{M.A.W.}}:
\batitle{{The Ursa Major Cluster of Galaxies. V. H I Rotation Curve Shapes and
  the Tully-Fisher Relations}}.
\bjtitle{\apj}
\bvolume{563},
\bfpage{694}--\blpage{715}
(\byear{2001})
{\href{https://arxiv.org/abs/astro-ph/0108225}{{astro-ph/0108225}}}.
\doiurl{10.1086/323887}
\end{barticle}
\endbibitem

%%% 25
\bibitem{2017A&A...597A..39M}
\begin{barticle}
\bauthor{\bsnm{{Michtchenko}}, \binits{T.A.}},
\bauthor{\bsnm{{Vieira}}, \binits{R.S.S.}},
\bauthor{\bsnm{{Barros}}, \binits{D.A.}},
\bauthor{\bsnm{{L{\'e}pine}}, \binits{J.R.D.}}:
\batitle{{Modelling resonances and orbital chaos in disk galaxies. Application
  to a Milky Way spiral model}}.
\bjtitle{\aap}
\bvolume{597},
\bfpage{39}
(\byear{2017})
{\href{https://arxiv.org/abs/1608.08991}{{arXiv:1608.08991}}}
{[astro-ph.GA]}.
\doiurl{10.1051/0004-6361/201628895}
\end{barticle}
\endbibitem

%%% 26
\bibitem{2012msma.book.....F}
\begin{bbook}
\bauthor{\bsnm{{Feigelson}}, \binits{E.D.}},
\bauthor{\bsnm{{Babu}}, \binits{G.J.}}:
\bbtitle{{Modern} {Statistical} {Methods} {for} {Astronomy}}.
\bpublisher{{Cambridge University Press, 2012, Cambridge, UK}}, \blocation{ }
(\byear{2012})
\end{bbook}
\endbibitem

%%% 27
\bibitem{2015IAUS..311...82S}
\begin{bchapter}
\bauthor{\bsnm{{Sick}}, \binits{J.}},
\bauthor{\bsnm{{Courteau}}, \binits{S.}},
\bauthor{\bsnm{{Cuillandre}}, \binits{J.-C.}},
\bauthor{\bsnm{{Dalcanton}}, \binits{J.}},
\bauthor{\bsnm{{de Jong}}, \binits{R.}},
\bauthor{\bsnm{{McDonald}}, \binits{M.}},
\bauthor{\bsnm{{Simard}}, \binits{D.}},
\bauthor{\bsnm{{Tully}}, \binits{R.B.}}:
\bctitle{{The Stellar Mass of M31 as inferred by the Andromeda Optical {\&}
  Infrared Disk Survey}}.
In: \beditor{\bsnm{{Cappellari}}, \binits{M.}},
\beditor{\bsnm{{Courteau}}, \binits{S.}} (eds.)
\bbtitle{Galaxy Masses as Constraints of Formation Models}.
\bsertitle{IAU Symposium},
vol. \bseriesno{311},
pp. \bfpage{82}--\blpage{85}
(\byear{2015}).
\doiurl{10.1017/S1743921315003440}
\end{bchapter}
\endbibitem

%%% 28
\bibitem{2010MNRAS.406..264W}
\begin{barticle}
\bauthor{\bsnm{{Watkins}}, \binits{L.L.}},
\bauthor{\bsnm{{Evans}}, \binits{N.W.}},
\bauthor{\bsnm{{An}}, \binits{J.H.}}:
\batitle{{The masses of the Milky Way and Andromeda galaxies}}.
\bjtitle{\mnras}
\bvolume{406},
\bfpage{264}--\blpage{278}
(\byear{2010})
{\href{https://arxiv.org/abs/1002.4565}{{arXiv:1002.4565}}}
{[astro-ph.GA]}.
\doiurl{10.1111/j.1365-2966.2010.16708.x}
\end{barticle}
\endbibitem

%%% 29
\bibitem{2008AJ....135...20E}
\begin{barticle}
\bauthor{\bsnm{{Erwin}}, \binits{P.}},
\bauthor{\bsnm{{Pohlen}}, \binits{M.}},
\bauthor{\bsnm{{Beckman}}, \binits{J.E.}}:
\batitle{{The Outer Disks of Early-Type Galaxies. I. Surface-Brightness
  Profiles of Barred Galaxies}}.
\bjtitle{\aj}
\bvolume{135},
\bfpage{20}--\blpage{54}
(\byear{2008})
{\href{https://arxiv.org/abs/0709.3505}{{arXiv:0709.3505}}}.
\doiurl{10.1088/0004-6256/135/1/20}
\end{barticle}
\endbibitem

%%% 30
\bibitem{2013AJ....146..104H}
\begin{barticle}
\bauthor{\bsnm{{Herrmann}}, \binits{K.A.}},
\bauthor{\bsnm{{Hunter}}, \binits{D.A.}},
\bauthor{\bsnm{{Elmegreen}}, \binits{B.G.}}:
\batitle{{Surface Brightness Profiles of Dwarf Galaxies. I. Profiles and
  Statistics}}.
\bjtitle{\aj}
\bvolume{146},
\bfpage{104}
(\byear{2013})
{\href{https://arxiv.org/abs/1309.0004}{{arXiv:1309.0004}}}.
\doiurl{10.1088/0004-6256/146/5/104}
\end{barticle}
\endbibitem

%%% 31
\bibitem{1969ApJ...155..393P}
\begin{barticle}
\bauthor{\bsnm{{Peebles}}, \binits{P.J.E.}}:
\batitle{{Origin of the Angular Momentum of Galaxies}}.
\bjtitle{\apj}
\bvolume{155},
\bfpage{393}
(\byear{1969}).
\doiurl{10.1086/149876}
\end{barticle}
\endbibitem

%%% 32
\bibitem{2018ApJ...863L..37M}
\begin{barticle}
\bauthor{\bsnm{{Michtchenko}}, \binits{T.A.}},
\bauthor{\bsnm{{L{\'e}pine}}, \binits{J.R.D.}},
\bauthor{\bsnm{{P{\'e}rez-Villegas}}, \binits{A.}},
\bauthor{\bsnm{{Vieira}}, \binits{R.S.S.}},
\bauthor{\bsnm{{Barros}}, \binits{D.A.}}:
\batitle{{On the Stellar Velocity Distribution in the Solar Neighborhood in
  Light of Gaia DR2}}.
\bjtitle{\apj}
\bvolume{863},
\bfpage{37}
(\byear{2018})
{\href{https://arxiv.org/abs/1808.01501}{{arXiv:1808.01501}}}
{[astro-ph.GA]}.
\doiurl{10.3847/2041-8213/aad804}
\end{barticle}
\endbibitem

%%% 33
\bibitem{1538-3881-137-3-3520}
\begin{barticle}
\bauthor{\bsnm{Griv}, \binits{E.}},
\bauthor{\bsnm{Gedalin}, \binits{M.}},
\bauthor{\bsnm{Eichler}, \binits{D.}}:
\batitle{The stellar velocity distribution in the solar neighborhood:
  Deviations from the schwarzschild distribution}.
\bjtitle{The Astronomical Journal}
\bvolume{137}(\bissue{3}),
\bfpage{3520}
(\byear{2009})
\end{barticle}
\endbibitem

%%% 34
\bibitem{1972.book.....K}
\begin{bbook}
\bauthor{\bsnm{{Kulsrud}}, \binits{R.M.}}:
\bbtitle{{Enhancement} {of} {Relaxation} {Processes} {by} {Collective}
  {Effects}}.
\bpublisher{Scott Tremaine. ISBN 978-0-691-13026-2.}, \blocation{ }
(\byear{1972})
\end{bbook}
\endbibitem

%%% 35
\bibitem{2014ApJ...786L..18L}
\begin{barticle}
\bauthor{\bsnm{{Luhman}}, \binits{K.L.}}:
\batitle{{Discovery of a \textasciitilde250 K Brown Dwarf at 2 pc from the
  Sun}}.
\bjtitle{\apj}
\bvolume{786}(\bissue{2}),
\bfpage{18}
(\byear{2014})
{\href{https://arxiv.org/abs/1404.6501}{{arXiv:1404.6501}}}
{[astro-ph.GA]}.
\doiurl{10.1088/2041-8205/786/2/L18}
\end{barticle}
\endbibitem

%%% 36
\bibitem{2018ApJ...866..131M}
\begin{barticle}
\bauthor{\bsnm{{Moro-Mart{\'{\i}}n}}, \binits{A.}}:
\batitle{{Origin of 'Oumuamua. I. An Ejected Protoplanetary Disk Object?}}
\bjtitle{\apj}
\bvolume{866},
\bfpage{131}
(\byear{2018})
{\href{https://arxiv.org/abs/1810.02148}{{arXiv:1810.02148}}}
{[astro-ph.EP]}.
\doiurl{10.3847/1538-4357/aadf34}
\end{barticle}
\endbibitem

%%% 37
\bibitem{1983AA...118...90W}
\begin{barticle}
\bauthor{\bsnm{{Weissman}}, \binits{P.R.}}:
\batitle{{The mass of the Oort cloud}}.
\bjtitle{\aap}
\bvolume{118}(\bissue{1}),
\bfpage{90}--\blpage{94}
(\byear{1983})
\end{barticle}
\endbibitem

%%% 38
\bibitem{1986EM&P...36..187M}
\begin{barticle}
\bauthor{\bsnm{{Mendis}}, \binits{D.A.}},
\bauthor{\bsnm{{Marconi}}, \binits{M.L.}}:
\batitle{{A note on the total mass of comets in the solar system}}.
\bjtitle{Earth Moon and Planets}
\bvolume{36},
\bfpage{187}--\blpage{190}
(\byear{1986}).
\doiurl{10.1007/BF00057610}
\end{barticle}
\endbibitem

%%% 39
\bibitem{1993AdSpR..13..121W}
\begin{barticle}
\bauthor{\bsnm{{Witte}}, \binits{M.}},
\bauthor{\bsnm{{Rosenbauer}}, \binits{H.}},
\bauthor{\bsnm{{Banaszkiewicz}}, \binits{M.}},
\bauthor{\bsnm{{Fahr}}, \binits{H.}}:
\batitle{{The ULYSSES neutral gas experiment - Determination of the velocity
  and temperature of the interstellar neutral helium}}.
\bjtitle{Advances in Space Research}
\bvolume{13},
\bfpage{121}--\blpage{130}
(\byear{1993}).
\doiurl{10.1016/0273-1177(93)90401-V}
\end{barticle}
\endbibitem

%%% 40
\bibitem{2007ApJ...663..866D}
\begin{barticle}
\bauthor{\bsnm{{Draine}}, \binits{B.T.}},
\bauthor{\bsnm{{Dale}}, \binits{D.A.}},
\bauthor{\bsnm{{Bendo}}, \binits{G.}},
\bauthor{\bsnm{{Gordon}}, \binits{K.D.}},
\bauthor{\bsnm{{Smith}}, \binits{J.D.T.}},
\bauthor{\bsnm{{Armus}}, \binits{L.}},
\bauthor{\bsnm{{Engelbracht}}, \binits{C.W.}},
\bauthor{\bsnm{{Helou}}, \binits{G.}},
\bauthor{\bsnm{{Kennicutt}}, \binits{R.C.} \bsuffix{Jr.}},
\bauthor{\bsnm{{Li}}, \binits{A.}},
\bauthor{\bsnm{{Roussel}}, \binits{H.}},
\bauthor{\bsnm{{Walter}}, \binits{F.}},
\bauthor{\bsnm{{Calzetti}}, \binits{D.}},
\bauthor{\bsnm{{Moustakas}}, \binits{J.}},
\bauthor{\bsnm{{Murphy}}, \binits{E.J.}},
\bauthor{\bsnm{{Rieke}}, \binits{G.H.}},
\bauthor{\bsnm{{Bot}}, \binits{C.}},
\bauthor{\bsnm{{Hollenbach}}, \binits{D.J.}},
\bauthor{\bsnm{{Sheth}}, \binits{K.}},
\bauthor{\bsnm{{Teplitz}}, \binits{H.I.}}:
\batitle{{Dust Masses, PAH Abundances, and Starlight Intensities in the SINGS
  Galaxy Sample}}.
\bjtitle{\apj}
\bvolume{663},
\bfpage{866}--\blpage{894}
(\byear{2007})
{\href{https://arxiv.org/abs/astro-ph/0703213}{{astro-ph/0703213}}}.
\doiurl{10.1086/518306}
\end{barticle}
\endbibitem

%%% 41
\bibitem{2000JGR...10510317M}
\begin{barticle}
\bauthor{\bsnm{{Mann}}, \binits{I.}},
\bauthor{\bsnm{{Kimura}}, \binits{H.}}:
\batitle{{Interstellar dust properties derived from mass density, mass
  distribution, and flux rates in the heliosphere}}.
\bjtitle{Journal of Geophysical Research}
\bvolume{105},
\bfpage{10317}--\blpage{10328}
(\byear{2000}).
\doiurl{10.1029/1999JA900404}
\end{barticle}
\endbibitem

%%% 42
\bibitem{2018ApJ...868...25B}
\begin{barticle}
\bauthor{\bsnm{{Bromley}}, \binits{B.C.}},
\bauthor{\bsnm{{Kenyon}}, \binits{S.J.}},
\bauthor{\bsnm{{Brown}}, \binits{W.R.}},
\bauthor{\bsnm{{Geller}}, \binits{M.J.}}:
\batitle{{Nearby High-speed Stars in Gaia DR2}}.
\bjtitle{\apj}
\bvolume{868}(\bissue{1}),
\bfpage{25}
(\byear{2018})
{\href{https://arxiv.org/abs/1808.02620}{{arXiv:1808.02620}}}
{[astro-ph.GA]}.
\doiurl{10.3847/1538-4357/aae83e}
\end{barticle}
\endbibitem

%%% 43
\bibitem{1988A&A...192..117V}
\begin{barticle}
\bauthor{\bsnm{{van der Kruit}}, \binits{P.C.}}:
\batitle{{The three-dimensional distribution of light and mass in disks of
  spiral galaxies}}.
\bjtitle{\aap}
\bvolume{192},
\bfpage{117}--\blpage{127}
(\byear{1988})
\end{barticle}
\endbibitem

%%% 44
\bibitem{1988asco.book.....H}
\begin{bbook}
\bauthor{\bsnm{{Harwit}}, \binits{M.}}:
\bbtitle{{Astrophysical} {Concepts}}.
\bpublisher{{Springer-Verlag, New York ISBN 0-387-96683-8}}, \blocation{ }
(\byear{1988})
\end{bbook}
\endbibitem

%%% 45
\bibitem{2020Natur.581..269N}
\begin{barticle}
\bauthor{\bsnm{{Neeleman}}, \binits{M.}},
\bauthor{\bsnm{{Prochaska}}, \binits{J.X.}},
\bauthor{\bsnm{{Kanekar}}, \binits{N.}},
\bauthor{\bsnm{{Rafelski}}, \binits{M.}}:
\batitle{{A cold, massive, rotating disk galaxy 1.5 billion years after the Big
  Bang}}.
\bjtitle{\nat}
\bvolume{581}(\bissue{7808}),
\bfpage{269}--\blpage{272}
(\byear{2020})
{\href{https://arxiv.org/abs/2005.09661}{{arXiv:2005.09661}}}
{[astro-ph.GA]}.
\doiurl{10.1038/s41586-020-2276-y}
\end{barticle}
\endbibitem

%%% 46
\bibitem{2007ApJ...660L.117F}
\begin{barticle}
\bauthor{\bsnm{{Frebel}}, \binits{A.}},
\bauthor{\bsnm{{Christlieb}}, \binits{N.}},
\bauthor{\bsnm{{Norris}}, \binits{J.E.}},
\bauthor{\bsnm{{Thom}}, \binits{C.}},
\bauthor{\bsnm{{Beers}}, \binits{T.C.}},
\bauthor{\bsnm{{Rhee}}, \binits{J.}}:
\batitle{{Discovery of HE 1523-0901, a Strongly r-Process-enhanced Metal-poor
  Star with Detected Uranium}}.
\bjtitle{\apj}
\bvolume{660}(\bissue{2}),
\bfpage{117}--\blpage{120}
(\byear{2007})
{\href{https://arxiv.org/abs/astro-ph/0703414}{{arXiv:astro-ph/0703414}}}
{[astro-ph]}.
\doiurl{10.1086/518122}
\end{barticle}
\endbibitem

%%% 47
\bibitem{2005A&A...440.1153D}
\begin{barticle}
\bauthor{\bsnm{{del Peloso}}, \binits{E.F.}},
\bauthor{\bsnm{{da Silva}}, \binits{L.}},
\bauthor{\bsnm{{Porto de Mello}}, \binits{G.F.}},
\bauthor{\bsnm{{Arany-Prado}}, \binits{L.I.}}:
\batitle{{The age of the Galactic thin disk from Th/Eu nucleocosmochronology.
  III. Extended sample}}.
\bjtitle{\aap}
\bvolume{440}(\bissue{3}),
\bfpage{1153}--\blpage{1159}
(\byear{2005})
{\href{https://arxiv.org/abs/astro-ph/0506458}{{arXiv:astro-ph/0506458}}}
{[astro-ph]}.
\doiurl{10.1051/0004-6361:20053307}
\end{barticle}
\endbibitem

%%% 48
\bibitem{2005A&A...442..929A}
\begin{barticle}
\bauthor{\bsnm{{Alcob{\'e}}}, \binits{S.}},
\bauthor{\bsnm{{Cubarsi}}, \binits{R.}}:
\batitle{{Disk populations from HIPPARCOS kinematic data. Discontinuities in
  the local velocity distribution}}.
\bjtitle{\aap}
\bvolume{442},
\bfpage{929}--\blpage{946}
(\byear{2005}).
\doiurl{10.1051/0004-6361:20053563}
\end{barticle}
\endbibitem

\end{thebibliography}
\end{document}